\documentclass[aip,jcp,amsmath,amssymb,floatfix,reprint,twocolumn]{revtex4-2}
\usepackage{color,soul}
\usepackage{graphicx}
\usepackage{amsmath, amsthm, amssymb,mathtools}
\usepackage{amsthm}
\usepackage{multirow}
\usepackage[normalem]{ulem}
\usepackage{soul}
\usepackage{hyperref}
\newcommand{\be}{\begin{equation}}
\newcommand{\ee}{\end{equation}}
\newcommand{\bea}{\begin{eqnarray}}
\newcommand{\eea}{\end{eqnarray}}
\makeatletter

\newcommand{\Rmnum}[1]{\expandafter\@slowromancap\romannumeral #1@}
\makeatother\usepackage{array, makecell} 

\begin{document}

%\title{Mpemba effect in a system of an overdamped particle in a double well potential: exact results and other properties}
\title{Mpemba effect in a Langevin system: population statistics, metastability and other exact results}

\author{Apurba Biswas}
\email{apurbab@imsc.res.in}
\affiliation{The Institute of Mathematical Sciences, C.I.T. Campus, Taramani, Chennai 600113, India}
\affiliation{Homi Bhabha National Institute, Training School Complex, Anushakti Nagar, Mumbai 400094, India}
\author{R. Rajesh} 
\email{rrajesh@imsc.res.in}
\affiliation{The Institute of Mathematical Sciences, C.I.T. Campus, Taramani, Chennai 600113, India}
\affiliation{Homi Bhabha National Institute, Training School Complex, Anushakti Nagar, Mumbai 400094, India}
\author{Arnab Pal} 
\email{arnabpal@imsc.res.in}
\affiliation{The Institute of Mathematical Sciences, C.I.T. Campus, Taramani, Chennai 600113, India}
\affiliation{Homi Bhabha National Institute, Training School Complex, Anushakti Nagar, Mumbai 400094, India}

%\date{\today}

\begin{abstract}
The Mpemba effect is a fingerprint of the anomalous relaxation phenomenon wherein an initially hotter system equilibrates faster than an initially colder system when both are quenched to the same low temperature. Experiments on a single colloidal particle trapped in a carefully shaped double well potential have demonstrated this effect recently [\textit{Nature \textbf{584}, 64 (2020)}]. In a similar vein, here, we consider a piece-wise linear double well potential that allows us to demonstrate the Mpemba effect using an exact analysis based on the spectral decomposition of the corresponding Fokker-Planck equation. We elucidate the role of the metastable states in the energy landscape as well as the initial population statistics of the particles in showcasing the Mpemba effect. Crucially, our findings indicate that neither the metastability nor the asymmetry in the potential is a necessary or a sufficient condition for the Mpemba effect to be observed.
\end{abstract}

\maketitle

\noindent

\section{{\label{Introduction}}Introduction}

The Mpemba effect refers to the faster equilibration of a hotter system compared to a colder system when both are quenched to a final temperature which is the lowest ~\cite{Mpemba_1969}. 
Initially studied in water~\cite{Mpemba_1969,Mirabedin-evporation-2017, vynnycky-convection:2015, katz2009hot, david-super-cooling-1995, zhang-hydrbond1-2014,tao-hydrogen-2017,Molecular_Dynamics_jin2015mechanisms, gijon2019paths}, the effect has now been established as a more general anomalous relaxation phenomenon.  There now exists a wide range of physical systems where  experimental evidences about the existence of the Mpemba effect have been reported. Examples include magnetic alloys~\cite{chaddah2010overtaking}, polylactides~\cite{Polylactide}, clathrate hydrates~\cite{paper:hydrates},  and colloidal systems~\cite{kumar2020exponentially,kumar2021anomalous,bechhoefer2021fresh}.

There has been a great deal of theoretical effort to demonstrate the Mpemba effect in spin 
systems~\cite{PhysRevLett.124.060602,Klich-2019,klich2018solution,das2021should,PhysRevE.104.044114,teza2021relaxation}, spin glasses~\cite{SpinGlassMpemba}, molecular gases in contact with a thermal reservoir~\cite{moleculargas,gonzalez2020mpemba,gonzalez2020anomalous,PhysRevE.104.064127}, Markovian systems with restricted phase space~\cite{Lu-raz:2017,PhysRevResearch.3.043160}, Langevin systems \cite{Walker_2021,Busiello_2021,lapolla2020faster,degunther2022anomalous,walker2022mpemba}, active systems~\cite{schwarzendahl2021anomalous}, quantum systems~\cite{PhysRevLett.127.060401,nava2019lindblad,chatterjee2023quantum}, systems with phase transitions~\cite{holtzman2022landau,das2021should,zhang2022theoretical,teza2022eigenvalue}, and granular systems~\cite{Lasanta-mpemba-1-2017,Torrente-rough-2019,mompo2020memory,PhysRevE.102.012906,biswas2021mpemba,biswas2022mpemba,megias2022mpemba,biswas2023measure}. 
Spanning across various physical systems, different causes have been attributed to the Mpemba effect although no unified consensus exists to the underlying reason. However, it turns out that in the analytically tractable kinetic state models or in the Langevin systems, the so-called ``multi-dimensional rugged energy landscape'' picture provides an effective description of the Mpemba effect. In particular, the presence of one or multiple metastable minima in the free energy can trap a system at a lower energy more effectively than the same at higher temperature, resulting in a faster relaxation of the hotter system.

More on the experimental side, the Mpemba effect was 
demonstrated by Kumar and Bechhoefer in a system of a colloidal particle diffusing in a confining double well quartic potential with linear slopes near the domain boundaries \cite{kumar2020exponentially}. It was shown that the asymmetry in the potential, which was realized by introducing different widths for the left- and the right- end domains, is a key factor for the Mpemba effect. As the asymmetry in the domain widths is being increased, even a stronger version of the Mpemba effect emerges where the relaxation is exponentially faster for a hotter system. Notwithstanding demonstrating this remarkable anomalous relaxation phenomena, there are a few yet fundamental frontiers that still remain open. 
For example, can asymmetry in the potential depths (in addition to the asymmetric domains) result in the Mpemba effect? Is there a necessary or a sufficient `asymmetry' condition on the nature/shape of the potential that can universally underpin the Mpemba effect? Another question that intrigues our mind along this line: Is a double well potential necessary to realize the Mpemba effect in Langevin systems? A recent study showed the Mpemba effect in a simple piecewise constant potential configuration where the minima of the potentials were set at neutral equilibrium \cite{Walker_2021} thus breaking down our general intuitions based on the rugged landscape, metastablity and dis-balanced statistics of the particles' population. 

% However, other possibilities of inducing the Mpemba effect through asymmetries in terms of different depths of the potential wells without changing the domain widths were not explored. Thus, it is not clear about what sort of asymmetry in the potential configuration is necessary or sufficient to induce the Mpemba effect. Moreover, it is not well understood whether a double well potential is necessary to realize the Mpemba effect in Langevin systems. Recently for a piecewise constant potential where the minima of the potentials have neutral equilibrium, it was shown that the Mpemba effect can be observed, questioning the need for a metastable state~\cite{Walker_2021}.  

\begin{figure}
\centering
\includegraphics[width= 0.8\columnwidth]{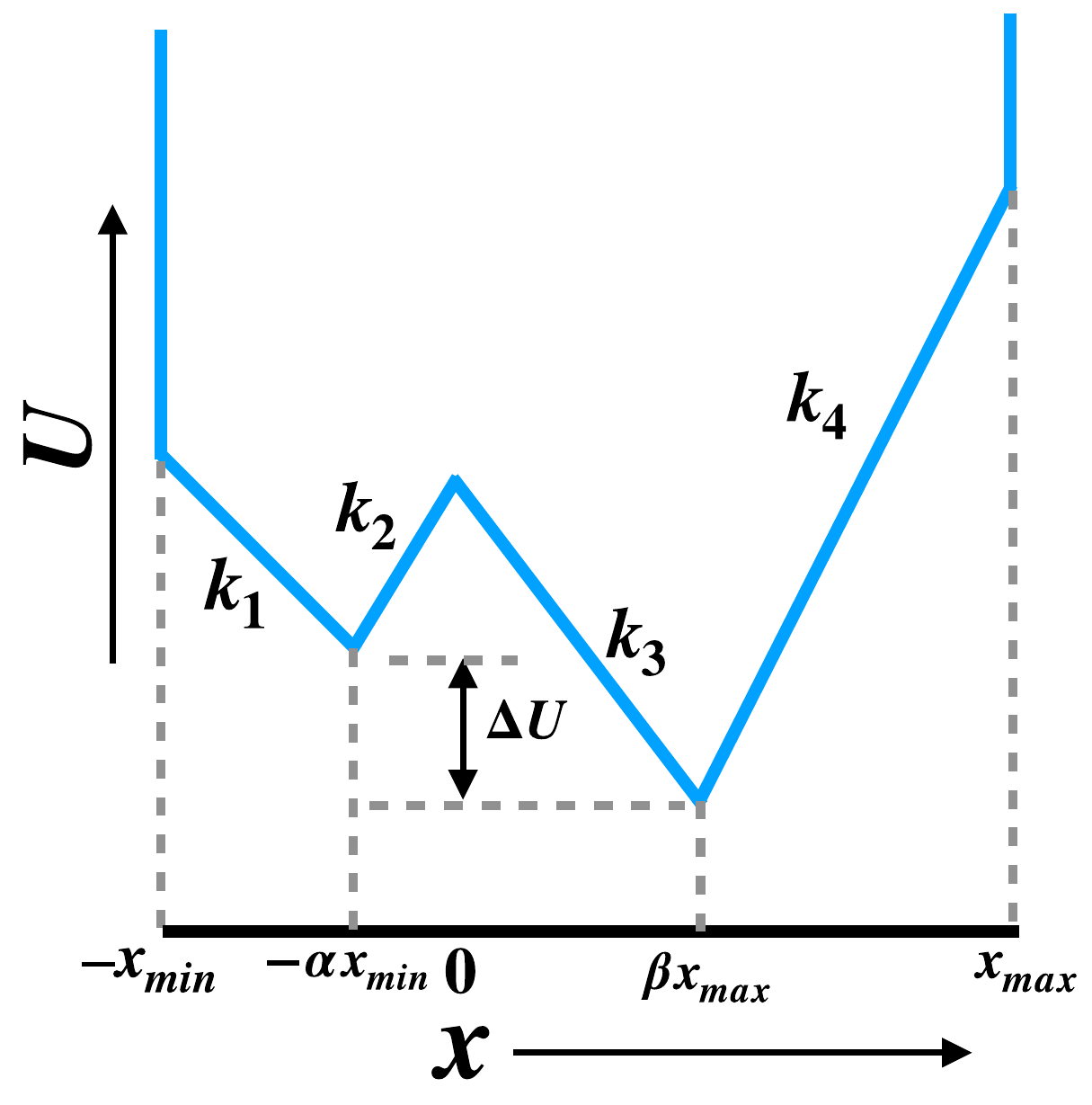} 
\caption{\label{potential shape}Schematic diagram of the piece-wise linear double well potential. The boundaries of the potential are situated at $-x_{min}$ and $x_{max}$. The two minima of the double well potential are located at $-\alpha x_{min}$ and $\beta x_{max}$, where $\alpha, \beta \in (0,1)$. The parameters $k_1$, $k_2$, $k_3$ and $k_4$ refer to the various slopes. $\Delta U$ depicts the difference between the depths of the two wells.}
\end{figure}

In this paper, we delve deeper into these questions by investigating an exactly solvable model. Similar to the experiment by Kumar and Bechhoefer~\cite{kumar2020exponentially}, we consider a system of an overdamped Brownian particle trapped in a double well potential. However, the potential is constructed in a piece-wise linear manner which makes the problem solvable. The major advantage besides its analytical tractability is that one can also gain deep insights from this canonical model by scanning several potential configurations. For example, in our set-up, the potential can be made asymmetric in several different ways: (a)  different widths for the left and right domains of the potential as discussed in the experiment by Kumar and Bechhoefer~\cite{kumar2020exponentially}, (b) same domain widths but asymmetric location of the potential minima, (c) different depths of the potential wells, and (d) different heights for the left, center and right edge of the potential (see Fig. \ref{potential shape}). The potential can also be transformed into a single well by taking certain limits. Solving for the time evolution of the probability distribution function using the method of eigenspectrum decomposition  of the corresponding Fokker-Planck equation~\cite{risken1996fokker,gardiner1985handbook}, we present a comprehensive analysis of the relaxation phenomena in terms of the eigenvalues and the distant functions for each of the above-mentioned cases. Our extensive analysis underpins the following observations: (i) asymmetry of domain widths is not a necessary condition for the existence of the Mpemba effect, (ii) asymmetry in the potential heights is not a sufficient condition for the Mpemba effect, and (iii) presence of the metastable states is neither a necessary nor a sufficient condition for this anomalous relaxation.

% In this paper, we  solve for  the time evolution of the probability density using the method of eigenspectrum decomposition  of the corresponding Fokker-Planck equation~\cite{risken1996fokker}. The potential can be made asymmetric in several different ways: (a)  different widths for the left and right domains of the potential as discussed in the experiment by Kumar and Bechhoefer~\cite{kumar2020exponentially}, (b) same domain widths but asymmetric placement of the potential minima, (c) different depths of the potential wells, and (d) for different heights for the left, center and right edge of the potential (see Fig. \ref{potential shape}). We investigate the presence of Mpemba effect for the above scenarios. In addition, we show that asymmetry of domain widths is not a necessary condition for the existence of the Mpemba effect. Also, we show that asymmetry in the potential is not a sufficient condition for the Mpemba effect. Finally, we  show through an example that the presence of a metastable state is not a necessary condition for the effect.   

The remainder of the paper is organized as follows. We 
describe our model system in Sec.~\ref{Model}. For this set-up, we sketch out the eigenspectrum decomposition method for solving the probability distribution function with suitable matching and boundary conditions. In Sec.~\ref{Distance function and the Mpemba effect}, we 
introduce the distance functions which measures the deviation of a transient state from an equilibrium state. Sec.~\ref{Modulation of the potential and population} discusses the role of population statistics of the Brownian particle across the double well potential landscape. We pinpoint the role of metastable states and discuss the variation in the population statistics of the Brownian particle as a result of the potential modulation. In Sec.~\ref{Illustration of the Mpemba effect}, we explore several `typical' and `atypical' configurations of the double well potential and illustrate the existence of the Mpemba effect. We provide two complete phase diagrams that demonstrate the possible configurations of the potential difference and hot-to-final temperature ratio where the Mpemba effect can be observed. Various intriguing facts are highlighted. In Sec.~\ref{Illustration of the Mpemba effect for single well}, we demonstrate the existence of the Mpemba effect in the absence of metastable states and discuss the intricate interplay between the population statistics and the initial kinetic energy of the Brownian particle that leads this effect. We conclude our paper in Section~\ref{Conclusion} with a brief summary and discussion.

\section{\label{Model}Model and general formalism}
We consider a colloidal particle diffusing in an asymmetric double well potential $\tilde{U}(x)$, as shown in Fig. \ref{potential shape}, in a thermal environment characterized by noise $\eta$ and damping $\gamma$. The mean and variance of the noise are
\be
\langle \eta(t) \rangle=0~~ \text{and}~~ \langle \eta(t) \eta(t') \rangle= 2 \gamma k_B T_b \delta (t-t'), \label{noise}
\ee
where $T_b$ is the temperature of the thermal bath and $k_B$ is the Boltzmann's constant. We consider the overdamped case where the damping $\gamma$ is large compared to the mass of the particle. Motion of the particle is then described by the overdamped Langevin equation
\be
\gamma \frac{dx}{dt}=-\frac{d\tilde{U}}{dx}+\eta(t). \label{langevin eqn}
\ee
The corresponding Fokker-Planck equation for the probability distribution function $p(x,t)$ reads \cite{risken1996fokker,morsch1979one}
\be
\frac{\partial p}{\partial t}=\frac{\partial}{\partial x}\Big[\frac{1}{\gamma}\frac{d \tilde{U}}{dx} p \Big]+ D\frac{\partial^2 p}{\partial x^2}, \label{FP eqn1}
 \ee
where $D$ is the diffusion coefficient and is related to the temperature via the Einstein's relation \cite{risken1996fokker}
 \be
 D=\frac{k_B T_b}{\gamma}.
 \ee
In here, we will solve $p(x,t)$ analytically for the given configuration of the potential. To this end, it will be first useful to review the formalism of eigenspectrum decomposition for solving the Fokker-Planck equation~(\ref{FP eqn1}) in the presence of a generic confining potential. 
 
 \subsection{Spectral decomposition \label{decomposition}}
 To layout the formalism we closely follow Risken \cite{risken1996fokker}. We start by normalizing the potential in terms of $k_BT_b$ so that
 \be
 U(x)=\frac{\tilde{U}(x)}{k_B T_b}. \label{potential and temp}
 \ee
 Thus, Eq.~(\ref{FP eqn1}) simplifies to the continuity equation
 \be
\frac{\partial p}{\partial t}=D\frac{\partial}{\partial x}\Big[\frac{d U}{dx} p \Big]+ D\frac{\partial^2 p}{\partial x^2}=-\frac{\partial J}{\partial x}, \label{FP eqn2}
 \ee
from where the probability current/flux can be identified as
 \be
 J(x)=
 -D\Big[\frac{dU}{dx}  + \frac{\partial }{\partial x} \Big]p=-De^{-U(x)} \frac{d}{dx} \Big[e^{U(x)} p \Big], \label{prob current}
 \ee
and the corresponding Fokker-Planck operator reads
 \be
 \mathcal{L}_{FP}=D\frac{\partial}{\partial x}\Big(\frac{dU}{dx} \Big)+D\frac{\partial^2 }{\partial x^2}. \label{FP operator}
 \ee
 The stationary solution of the Fokker-Planck Eq.~(\ref{FP eqn2}) for the probability density is given by the Boltzmann distribution at temperature $T_b$
 \be
 \pi(x,T_b)= \frac{e^{-U(x)}}{\mathcal{Z}},
 \ee
 where $\mathcal{Z}=\int e^{-U(x)} dx$ is the partition function. Note that the Fokker-Planck operator in Eq.~(\ref{FP operator}) is not self-adjoint. A simple transformation leads to its  self-adjoint form $\mathcal{L}$ where
 \bea
 \mathcal{L}=e^{U(x)} \mathcal{L}_{FP} e^{-U(x)}=D\frac{\partial^2}{\partial x^2}-V(x), \label{adjoint FP operator}
 \eea
and
\bea
 V(x)=\frac{D}{4}\Big(\frac{dU}{dx}\Big)^2- \frac{D}{2}\frac{d^2 U}{dx^2},
 \eea
 is now the effective potential.  Thus the original problem is now reduced to analyzing the following eigenvalue problem
\bea
\mathcal{L} \psi_n =\lambda_n \psi_n, \label{FP eigenvalue1}
\eea
where $ \psi_n$ are the eigenfunctions of the self-adjoint Fokker-Planck operator $\mathcal{L}$ corresponding to the eigenvalue $\lambda_n$. Denoting the eigenvectors of the Fokker-Planck operator $\mathcal{L}_{FP}$ as $\phi_n(x)$ and noting that both of them have the same eigenvalues, one can write \cite{risken1996fokker}
\be
\psi_n(x)=e^{\frac{U(x)}{2}} \phi_n(x). \label{wavefunction relation}
\ee
The eigenvalues $\lambda_n$ follow the order: $\lambda_1=0>\lambda_2>\lambda_3 \ldots$, where $\lambda_1=0$ corresponds to the stationary distribution for a bath temperature, $T_b$. The first eigenvector corresponding to $\lambda_1=0$ is given by $\psi_1(x)=e^{-U(x)/2}/\sqrt{\mathcal{Z}(T_b)}$. 

Given the initial probability condition $p(x',0)$, the probability distribution function $p(x,t)$ can be obtained as
\be
p(x,t)=\int  W(x,t|x',0) p(x',0) dx', \label{eq p}
\ee
where the transition probability or the propagator $W(x,t|x',0)$ of the Fokker-Planck equation can be written in terms of the eigenfunctions and eigenvalues (see \cite{risken1996fokker,morsch1979one})
\begin{align}
    W(x,t|x',0) &= e^{\mathcal{L}_{FP}t}~ \delta(x-x') \nonumber \\
    &= e^{-\frac{U(x)}{2}+\frac{U(x')}{2}} \sum_n e^{\lambda_n t}    \psi_n(x) \psi^{*}_n(x').
\end{align}
Substituting the transition probability into Eq. (\ref{eq p}), one finds
\be
p(x,t)=\int dx' e^{-\frac{U(x)}{2}+\frac{U(x')}{2}} \sum_n e^{\lambda_n t}    \psi_n(x) \psi^{*}_n(x') p(x',0). \label{prob density}
\ee
Since $\lambda_1=0$, we can rewrite Eq.~(\ref{prob density}) as follows
\be
p(x,t)=\frac{e^{-U(x)}}{\mathcal{Z}(T_b)}+\sum_{n\geq 2} a_n e^{\frac{-U(x)}{2}} \psi_n(x) e^{-|\lambda_n| t}, \label{prob density solution}
\ee
where 
\be
a_n=\int dx' ~p(x',0) ~e^{\frac{U(x')}{2}} ~\psi^{*}_n(x'). \label{a2 coeff}
\ee
At large times, since $\lambda_2>\lambda_3$, to leading order, we obtain
\be
p(x,t)\simeq \frac{e^{-U(x)}}{\mathcal{Z}(T_b)}+  a_2 e^{\frac{-U(x)}{2}} \psi_2(x) e^{-|\lambda_2| t},~t \gg \frac{1}{|\lambda_3|}. \label{prob density approx}
\ee
The equation above is central to further analysis of the relaxation properties for the particle in the potential $\Tilde{U}(x)$.

\subsection{\label{Form of the potential}Shape of the potential}
The form of the potential well is crucial to the observation of the Mpemba effect as was demonstrated in the experiment \cite{kumar2020exponentially}. In there, $\tilde{U}(x)$ is considered to be a double well quartic potential with linear slopes near its boundaries or domain walls. Furthermore, the potential is confined in an asymmetric domain and it was shown that the asymmetry in the widths of the left and right domains about the origin can lead to  the Mpemba effect \cite{kumar2020exponentially}.  

Likewise, we consider a double well potential which is piece-wise linear. In contrast to the quartic double well potential, this problem is exactly solvable as will be evident below. The boundaries of the well are situated at $(-x_{min}, x_{max})$. For simplicity, we set $k_BT_b=1$. The potential in Fig.~\ref{potential shape} can be quantified in the following way
\begin{widetext}
\be
U(x)=
\begin{cases}
-k_1 x, & -x_{min} < x < -\alpha x_{min} \\
k_2 x +\alpha (k_1 + k_2) x_{min}, & -\alpha x_{min} < x < 0 \\
-k_3 x +\alpha (k_1 + k_2) x_{min}, & 0 < x < \beta x_{max} \\
k_4 x +[\alpha (k_1 + k_2)- \beta (k_3 + k_4)] x_{max}, & \beta x_{max} < x < x_{max},  \label{potential form}
\end{cases}
\ee
\end{widetext}
where $k_1$, $k_2$, $k_3$ and $k_4$ are slope constants that play a crucial role in designating the potential various shapes and the two constants $\alpha, \beta \in (0,1)$.

The asymmetry in the shape of the potential can be introduced through various parameters such as different domain widths about the origin, different positions of the two wells about the origin or due to the different depths of the potential wells. However, it turns out that the different heights of the two wells is a key factor to the observation of the Mpemba effect in contrary to the result shown in Ref.~\cite{kumar2020exponentially}. This potential set-up provides an amenable physical interpretation for underlying cause in such systems as will be discussed and illustrated in Sec.~\ref{Modulation of the potential and population}.

\subsection{\label{Jump conditions}Jump conditions}

The potential in Eq.~(\ref{potential form}) is not differentiable at $x=-\alpha x_{min}$, $0$ and $\beta x_{max}$, and  diverges at the boundaries $x=-x_{min}$ and  $x=x_{max}$. Let $x_-$ and $x_+$ denote the points just to the left and right of boundary of a linear segment. For the choice of potential $U(x_+)=U(x_-)$ while $U'(x_+)\neq U'(x_-)$. Across a boundary, both the probability currents are equal, i.e., $J(x_+,t)=J(x_-,t)$, as well as the probabilities are equal. Thus, from Eq.~(\ref{prob current}), we obtain
\bea
&&-U'(x_+) p(x_+,t)-\frac{\partial p(x_+,t)}{\partial x} 
=\nonumber\\
&&-U'(x_-) p(x_-,t)-\frac{\partial p(x_-,t)}{\partial x}. \label{current density continuity}\\
&& p(x_+,t)= p(x_-,t). \label{current density continuity 2}
\eea
The jump conditions in Eqs.~(\ref{current density continuity}) and (\ref{current density continuity 2}) are satisfied by each of the eigenfunctions, and hence from Eq.~(\ref{prob density}),  we have
\bea
&&\psi'_n(x_+) + \frac{U'(x_+) \psi_n(x_+) }{2} 
=\psi'_n(x_-) + \frac{U'(x_-) \psi_n(x_-)}{2}  \label{jump 2},\\
&& \psi_n(x_+) = \psi_n(x_-). \label{jump 1}
\eea
At the boundaries, the potential diverges. This implies that the probability current must vanish and it leads to the following condition in terms of the eigenfunctions
\bea
\psi_n'(x)+\frac{U'(x)}{2} \psi_n(x)=0,~ \text{at}~x=-x_{min},~x_{max}. \label{boundary condition}
\eea
The jump conditions [Eqs.~(\ref{jump 1}), (\ref{jump 2}) and (\ref{boundary condition})] are utilized to solve the eigenspectrum of the Fokker-Planck operator $\mathcal{L}$ [see Eq.~(\ref{FP eigenvalue1})] as discussed in the next section.

\subsection{\label{eigenspectrum}Eigenspectrum analysis}
We have the task to solve the following eigenvalue problem \bea
\mathcal{L} \psi_n =-|\lambda_n| \psi_n, \label{ev eqn}
\eea
where $ \psi_n$ are the eigenfunctions of the self-adjoint Fokker-Planck operator $\mathcal{L}$ [see Eq.~(\ref{adjoint FP operator})] corresponding to the eigenvalue $\lambda_n$.  We  solve this equation separately in each of the four domains of the potential $U(x)$, characterized by slopes $k_1$, $k_2$, $k_3$, and $k_4$. This will lead to eight constants of integration which will be determined by the jump conditions at the boundaries of the regions, leading to a transcendental equation for the eigenvalue. Each of these cases is discussed in below.

\subsubsection{ Region~I: $-x_{min}<x<-\alpha x_{min}$}
In this given region, we have $U'(x)=-k_1$. Then, Eq.~(\ref{ev eqn}) takes the form:
\be
\frac{d^2 \psi^{\Rmnum{1}}_{n}}{dx^2}+\Big(\frac{\lambda_n}{D} - \frac{k^2_1}{4} \Big) \psi^{\Rmnum{1}}_n=0,
\ee
which has the solution
\be
\psi^{\Rmnum{1}}_n(x)=A_n \sin(m_{1n} x) + B_n \cos(m_{1n} x),
\label{eq:psi1}
\ee
where $A_n$, $B_n$ are constants  and 
\be
m_{1n}=\sqrt{\frac{\lambda_n}{D} - \frac{k^2_1}{4}}.
\ee

The solutions for the eigenfunctions in the other regimes are similar, but with different constants. We list them below. 

\subsubsection{Region~II: $-\alpha x_{min}<x<0$}
Here, we have $U'(x)=k_2$ and the solution for the eigenfunction is
\bea
\psi^{\Rmnum{2}}_n(x)=C_n \sin(m_{2n} x) + D_n \cos(m_{2n} x),
\label{eq:psi2}
\eea
where 
\bea
m_{2n}=\sqrt{\frac{\lambda_n}{D} - \frac{k^2_2}{4}}.
\eea

\subsubsection{Region~III: $0<x<\beta x_{max}$}
In this case, we have $U'(x)=-k_3$ and the solution reads
\bea
\psi^{\Rmnum{3}}_n(x)=E_n \sin(m_{3n} x) + F_n \cos(m_{3n} x),
\label{eq:psi3}
\eea
where 
\bea
m_{3n}=\sqrt{\frac{\lambda_n}{D} - \frac{k^2_3}{4}}.
\eea

\subsubsection{Region~IV: $\beta x_{max}<x<x_{max}$}
In here, we have $U'(x)=k_4$ and the solution for the eigenfunction is given by
\bea
\psi^{\Rmnum{4}}_n(x)=G_n \sin(m_{4n} x) + H_n \cos(m_{4n} x),
\label{eq:psi4}
\eea
where 
\be
m_{4n}=\sqrt{\frac{\lambda_n}{D} - \frac{k^2_4}{4}}.
\ee

\subsection{Boundary and matching conditions}
We now determine the different constants using the matching and boundary conditions. While the boundary conditions [see Eq.~(\ref{boundary condition})] are associated with the divergence of the potential at the boundaries leading to the vanishing probability current,  the matching conditions [see Eqs.~(\ref{jump 2}) and (\ref{jump 1})] arise at the boundaries of the potential domains due to the continuity of the probability current. 

\subsubsection{Boundary condition at $x=x_{max}$:}
The infinite jump in potential at $x=x_{max}$ leads to the vanishing probability current. It is given by the boundary condition [Eq.~(\ref{boundary condition})] which in terms of the eigenfunction $\psi^{\Rmnum{4}}_n$ reads
\be
\psi^{\Rmnum{4} \prime}_n(x_{max})+\frac{U'_4(x_{max})}{2} \psi^{\Rmnum{4}}_n(x_{max})=0.
\ee
Substituting for $\psi^{\Rmnum{4}}_n$ from Eq.~(\ref{eq:psi4}), we obtain 
\bea
G_n&=&-\nu_{4n} H_n,\\
\nu_{4n}&=&\frac{\frac{k_{4}}{2} \cos(m_{4n} x_{max}) -  m_{4n} \sin(m_{4n} x_{max}) }{\frac{k_{4}}{2} \sin(m_{4n} x_{max}) +  m_{4n} \cos(m_{4n} x_{max}) }.
\eea
Thus,
\bea
\psi^{\Rmnum{4}}_n(x)=H_n \left[ \cos(m_{4n} x) - \nu_{4n}\sin(m_{4n} x) \right].
\eea

\subsubsection{Boundary condition at $x=-x_{min}$:}
% $\boldsymbol{(f)}$ $\textbf{Boundary condition at $-x_{min}$}$:
Similar to the boundary condition at $x=x_{max}$, there is a divergence in the potential at $x=-x_{min}$ in the form of an infinite jump. The boundary condition in Eq.~(\ref{boundary condition}),  in terms of the 
 eigenfunctions $\psi^{\Rmnum{1}}_n$, is then given by
\be
\psi^{\Rmnum{1}\prime}_n(-x_{min})+\frac{U'_1(-x_{min})}{2} \psi^{\Rmnum{1}}_n(-x_{min})=0.
\ee
Substituting for $\psi^{\Rmnum{1}}_n$ from Eq.~(\ref{eq:psi1}), we obtain
\bea
A_n&=&\nu_{1n} B_n,\\
\nu_{1n}&=&\frac{\frac{k_{1}}{2} \cos(m_{1n} x_{min}) -  m_{1n} \sin(m_{1n} x_{min}) }{\frac{k_{1}}{2} \sin(m_{1n} x_{min}) +  m_{1n} \cos(m_{1n} x_{min}) }.
\eea
Thus,
\be
\psi^{\Rmnum{1}}_n(x)=B_n \left[ \cos(m_{4n} x) + \nu_{1n}\sin(m_{4n} x) \right].
\ee
We now use the jump conditions associated with the continuity of the probability current [see Eqs.~(\ref{jump 2})  and (\ref{jump 1})] across the boundaries of the potential domains at $x=-\alpha x_{min}, 0$ and  $x=\beta x_{max}$. They are given by the following three matching conditions. \\

\subsubsection{Matching condition at $x=-\alpha x_{min}$}
At $x=-\alpha x_{min}$, the eigenfunctions $\psi^{\Rmnum{1}}_n(x)$ and $\psi^{\Rmnum{2}}_n(x)$ satisfy the matching conditions given by Eqs.~(\ref{jump 2}) and (\ref{jump 1}) which simplifies to 
\begin{widetext}
\begin{align}
&C_n \Big[m_2 \cos(m_2 \alpha x_{min})-\frac{k_2}{2} \sin(m_2 \alpha x_{min}) \Big]
+ D_n \Big[m_2 \sin(m_2 \alpha x_{min})+\frac{k_2}{2} \cos(m_2 \alpha x_{min}) \Big] \nonumber \\
&=B_n \Big[ (m_1 \nu_{1n} -\frac{k_1}{2})\cos(m_1 \alpha x_{min}) 
 + (m_1 + \frac{\nu_{1n} k_1}{2}) \sin(m_1 \alpha x_{min}) \Big], \label{coeff 2}\\
\text{and}\nonumber \\
&- C_n \sin(m_2 \alpha x_{min}) + D_n \cos(m_2 \alpha x_{min})   = B_n [\cos(m_1 \alpha x_{min})-\nu_{1n} \sin(m_1 \alpha x_{min})], \label{coeff 1}
\end{align}
\end{widetext}
respectively. The coefficients $C_n$ and $D_n$ are solved using Eqs.~(\ref{coeff 1}) and (\ref{coeff 2}) in terms of $B_n$  and the expressions are given in Eqs.~(\ref{C1}) and (\ref{D1}) of Appendix~\ref{appendix 1}.

\subsubsection{Matching condition at $x=0$}

 Similar to the above, the eigenfunctions $\psi^{\Rmnum{2}}_n(x)$ and $\psi^{\Rmnum{3}}_n(x)$
 satisfy the matching conditions given by Eqs.~(\ref{jump 2}) and (\ref{jump 1}) for the continuity of the probability current at $x=0$ which simplifies to 
\bea
E_n&=&\frac{m_2}{m_3}C_n + \frac{k_2+k_3}{2m_3}D_n, \\
F_n&=&D_n.
\eea

\subsubsection{Matching condition at $x=\beta x_{max}$:}

At $x=\beta x_{max}$, the matching conditions given by Eqs.~(\ref{jump 2}) and (\ref{jump 1}) is satisfied by the eigenfunctions $\psi^{\Rmnum{3}}_n(x)$ and $\psi^{\Rmnum{4}}_n(x)$,  which  simplifies to
\bea
\begin{split}
&C_n \frac{m_2}{m_3} \sin(m_3 \beta x_{max}) \\
&+ D_n \Big(\frac{k_2+k_3}{2m_3} \sin(m_3 \beta x_{max}) + \cos(m_3 \beta x_{max}) \Big)  \\
&=H_n \Big[\cos(m_4 \beta x_{max}) - \nu_{4n} \sin(m_4 \beta x_{max}) \Big],\label{coeff 3}
\end{split}
\eea
and
\bea
\begin{split}
&C_n \Big[m_2 \cos(m_3 \beta x_{max})-\frac{k_3}{2}\frac{m_2}{m_3} \sin(m_3 \beta x_{max}) \Big]  \\
& +D_n \Big[\frac{k_2}{2}\cos(m_3\beta x_{max}) \\
&-\Big(m_3 + \frac{k_3(k_2+k_3)}{4 m_3} \Big) \sin(m_3\beta x_{max}) \Big]  \\
&=H_n \Big[ \Big(\frac{k_4}{2}-\nu_{4n} m_4 \Big) \cos(m_4 \beta x_{max})  \\
&- \Big(\frac{\nu_{4n}k_4}{2}+m_4 \Big)\sin(m_4 \beta x_{max}) \Big], \label{coeff 4}
\end{split}
\eea
respectively. The coefficients $C_n$ and $D_n$ are solved in terms of $H_n$ using Eqs.~(\ref{coeff 3}) and (\ref{coeff 4})  and the expressions are given in Eqs.~(\ref{C2}) and (\ref{D2}) of Appendix~\ref{appendix 1}. Now, we consider the ratios of Eqs.~(\ref{C1}), (\ref{D1}), (\ref{C2}) and (\ref{D2}) that form a transcendental equation to solve for the eigenvalues $\lambda_n$. Thus, solving for the eigenvalues $\lambda_n$ in turn helps to find the constants $A_n$, $B_n$, $C_n$, $D_n$, $E_n$, $F_n$ and $H_n$.

\section{\label{Distance function and the Mpemba effect}Distance function and the Mpemba effect}
How to quantify the Mpemba effect as an anomalous relaxation phenomena? To see this, let us consider two systems: first one $P$, initially equilibrated at temperature $T_h$ and second one $Q$, initially equilibrated at temperature $T_c$ where  $T_h > T_c$.  These initial equilibrium distributions are denoted by $\pi(T_h)$ and $\pi(T_c)$ respectively. Now imagine that both $P$ and $Q$ are quenched at once to a common bath temperature, $T_b$, where $T_h > T_c>T_b$. Eventually, both of them will equilibrate to the common distribution  $\pi(T_b)$ given long enough time. The Mpemba effect is said to exist if $P$ equilibrates faster than $Q$ during the transient/relaxation process. 

To quantify this relaxation process, let us now define the \textit{distance from equilibrium function}, $D[p(t),\pi(T_b)]$ which measures the instantaneous distance of a distribution $p(x,t)$ from the final equilibrium Boltzmann distribution, $\pi(T_b)$. It has been argued (see \cite{Lu-raz:2017} and others) that the Mpemba effect is independent of $D[p(t),\pi(T_b)]$ provided that the distance measure obeys the following properties: (a) If  $T_h>T_c>T_b$, then the distance from equilibrium function should follow the order $D[\pi(T_h), \pi(T_b)]>D[\pi(T_c), \pi(T_b)]$, (b) $D[p(t), \pi(T_b)]$ should be a monotonically non-increasing function of time, and (c) $D[p(t), \pi(T_b)]$ should be a convex function of $p(x,t)$. 

Notably, there are many well-adapted measures that exist in the literature namely the entropic distance, $L_1$ or norm distance and the Kullback-Leibler (KL) divergence~\cite{Lu-raz:2017,kumar2020exponentially,Busiello_2021}. Thus, as a working definition, if one has $D[\pi(T_h), \pi(T_b)]>D[\pi(T_c),\pi(T_b)]$ initially for $T_h > T_c$ followed by $D[p^h(t),\pi(T_b)]<D[p^c(t),\pi(T_b)]$ at a later time, we will state that  the Mpemba effect exists. 

% We define the Mpemba effect as follows. By definition, since $T_h > T_c$, initially $D[\pi(T_h), \pi(T_b)]>D[\pi(T_c),\pi(T_b)]$. If at late times,  $D[p^h(t),\pi(T_b)]<D[p^c(t),\pi(T_b)]$, then system $P$ will equilibrate faster, and we will say that the Mpemba effect exists. 

In the rest of the article, we will use the ``$L_1$ or norm" measure for the distance from equilibrium function. More precisely, this is defined as
\bea
\begin{split}
D[p(t),\pi(T_b)] &\equiv L_1(t)=\int dx |p(x,t)-\pi(x,T_b)|. \label{L2 measure-def}
\end{split}
\eea
Now substituting the form of $p(x,t)$ from Eq. (\ref{prob density approx}) into the above equation, we find
\begin{align}
    D[p(t),\pi(T_b)]= \sum_{n\geq 2} |a_n e^{\frac{-U(x)}{2}} \psi_n(x)| e^{-|\lambda_n| t}.
     \label{L2 measure}
\end{align}
 The condition for the Mpemba effect, as mentioned above, now boils down to  
 \begin{align}
|a_2(T_c)| \equiv |a^c_2|>|a^h_2| \equiv |a_2(T_h)|~.     
 \end{align}
  The condition demands that $|a_2(T)|$ should have a non-monotonic behavior with the increase in temperature. Note that the coefficient $a_2$ is calculated using Eq.~(\ref{a2 coeff}) and is a function only of the initial temperature and the bath temperature. Also, $a_2(T)$ is zero at the final temperature $T=T_b$ since the eigenvectors are orthonormal.

\section{\label{Modulation of the potential and population}Modulation of the potential and population -- connection to the experiments}
Following the colloidal experiment by Kumar and Bechhoefer, we learnt that the asymmetry in the shape of the double well potential plays an important role to the Mpemba effect. In particular, it was shown that there is no such effect if the asymmetry in the width of the left and right domains of the potential vanishes \cite{kumar2020exponentially}. In this section, we aim to revisit these limits from our model system by suitably changing the potential barrier.

\begin{figure}
\centering
\includegraphics[width= \columnwidth]{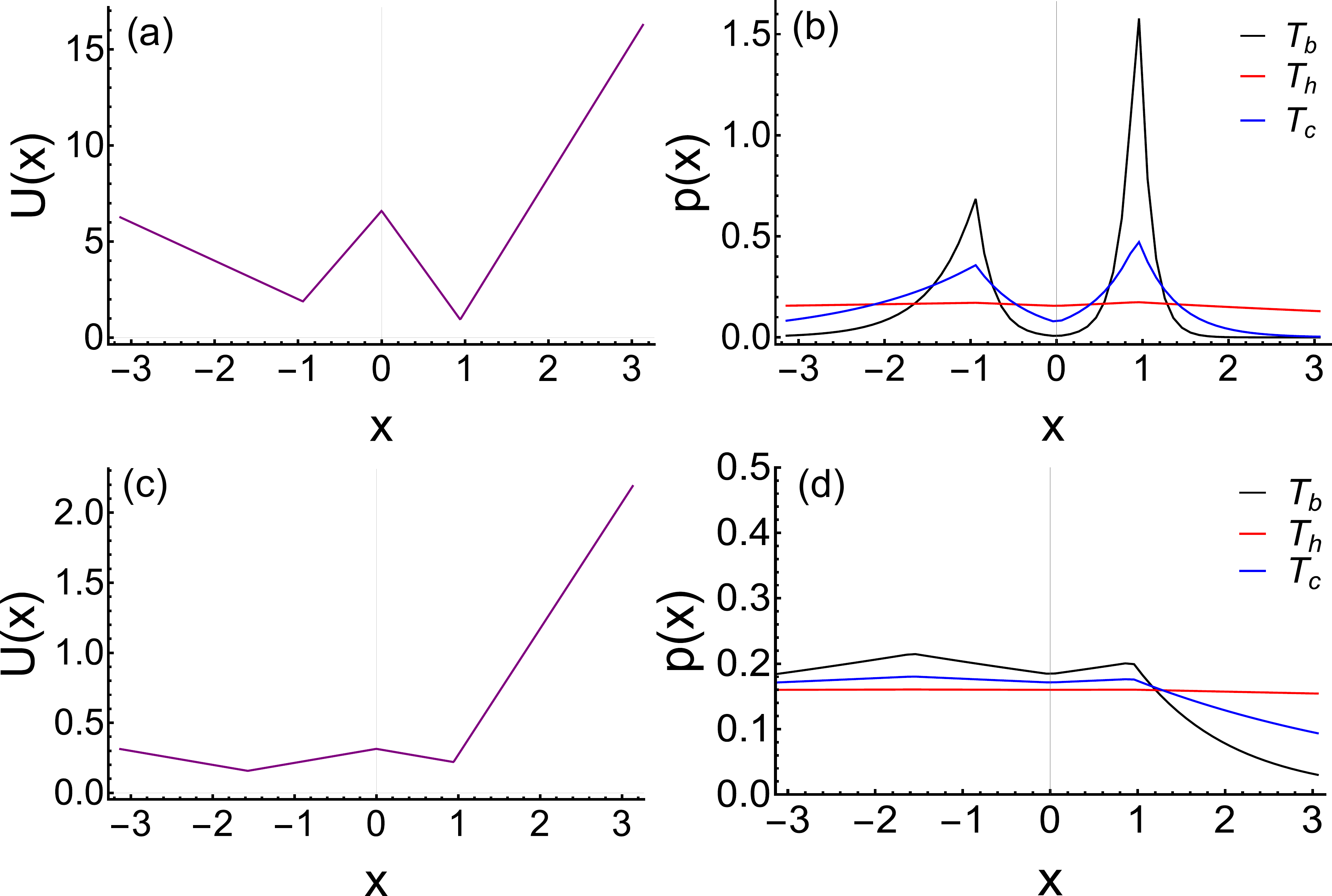} 
\caption{\label{modulation}Modulation of the potential and its effect on the population distribution of the Brownian particle in the two wells. Panel (a) and (c) corresponds to different configurations of the potential well. The parameters of the potential in (a) are chosen to be $\alpha=0.3$, $\beta=0.3$, $x_{min}=\pi$, $x_{max}=\pi$, $D=1$, $k_1=2$, $k_2=5$, $k_3=6$ and $k_4=7$ and in (c) $\alpha=0.5$, $\beta=0.3$, $|x_{min}|=\pi$, $x_{max}=\pi$, $D=1$, $k_1=0.1$, $k_2=0.1$, $k_3=0.1$ and $k_4=0.9$. Panels (b) and (d) depict the population distribution corresponding to the external potentials in (a) and (c) respectively for the initial temperatures $T_h$ (red), $T_c$ (blue) and final temperature $T_b$ (black) with $T_h>T_c>T_b$.}
\end{figure}

To this end, let us turn our attention to  Figs. ~\ref{modulation}(a) and \ref{modulation}(c) which show two different configurations for the potential barrier. The modulation of the potential barrier leads to a rearrangement in the population of the Brownian particle between the two wells for the two different temperatures as shown in   Fig.~\ref{modulation}(b) and \ref{modulation}(d).

In Fig.~\ref{modulation}(a), we consider a configuration of a potential with a considerable  potential barrier between the two minima. The corresponding population distribution of the Brownian particle for the temperatures $T_h$, $T_c$ and $T_b$, i.e., for the hot, cold and the bath temperatures respectively is shown in Fig.~\ref{modulation}(b). The initially colder system is more populated in the lowest well compared to the initially hotter system. However, there is also a considerable amount of population distributed in the metastable state for the colder system, which is not the case for the initially hotter system whose population distribution is nearly uniform i.e., it can not really `see' the metastable state. As a result, post quenching, the population distribution of the colder system takes a significant amount of time to rearrange and eventually relax to the lowest energy well from the metastable state. On the other hand, the initially hotter system ends with a higher population in the lowest well due to its fast relaxation. This feature grants an advantage to the initially hotter system over the colder one, and the Mpemba effect is observed.

Next, we consider the potential  shown in Fig.~\ref{modulation}(c), where the potential barrier between the two minima is almost diminishing thus creating a flat barrier between the wells. In this case, the population distribution at the bath temperature, $T_b$ which corresponds to the final equilibrium state, is almost equally populated between the two wells. Moreover, not much difference can be seen in the population distribution of the initially hot and the cold system. In other words, the `hindrance' due to the metastable state in the relaxation to the equilibrium state is absent. 
Owing to this, the relaxation process is similar for the both initially hot and the cold system. The initially colder system (having distribution closer to the final equilibrium state) relaxes faster compared to the initially hot system and hence, no Mpemba like effect is observed.

The above physical scenarios naturally set the stage to make the connection with the experiment \cite{kumar2020exponentially}. In particular, the asymmetry in the widths of the left and right domains of the confined potential in the experiment plays an analogous role to a finite barrier height between the wells in our model set-up. As we have shown that this configuration leads to the Mpemba effect similar to the
asymmetric domain for the left and the right well of the potential in the experiment. 

On the other hand, the symmetric double well potential configuration in the experiment with equal widths for the left and right domains is analogous to our second case with almost a flat barrier between the two wells of the potential [see Figs.~\ref{modulation}(c) and (d)]. It is because the symmetric potential configuration has the population of the particle almost similarly distributed between the two wells of the potential for any temperatures eliminating the effect of the presence of any metastable state. As a result, the relaxation dynamics from an initial equilibrium distribution to the final equilibrium are similar for any temperature, and the initially cold system having an initial temperature closer to the final equilibrium state relaxes faster. Hence, no Mpemba effect can be seen. These two possible configurations thus draw physical similarities between the experiment and our system.

% It is also the case with our chosen configuration of the potential with a flat barrier between the two wells of the potential, thus drawing similarity between the cases of symmetric potential configuration of the experiment and our case of the potential configuration with a flat barrier height.

%On the other hand, the symmetric double well potential (equal left and right potential domain widths) has the population of the Brownian particle equally distributed between the two wells for any temperature and in that case, there is no existence of the intermediate metastable state. As a result, the relaxation dynamics is similar for any temperature and the system having initial temperature closer to the final equilibrium state relaxes faster. Hence, there is no Mpemba effect for the symmetric potential well. \textcolor{red}{Apurba, please complete this part by stating why the symmetric part is equivalent to our second case here.}

\section{\label{Illustration of the Mpemba effect}Mpemba effect in double well potential}
In this section, we showcase several key configurations of the double well potential that can lead to the Mpemba effect. These results are analyzed based on the generic criterion for the Mpemba effect as described in Sec.~\ref{Distance function and the Mpemba effect}. 
% As discussed in Sec.~\ref{Modulation of the potential and population}, the configuration of the confining potential plays a key role behind the existence of the Mpemba effect. In the experiment in Ref.~\cite{kumar2020exponentially} it was shown that  different domain widths lead to the Mpemba effect. 
The methodology we use is as follows. Given a configuration of the potential, we solve the eigenvalue Eq.~(\ref{ev eqn}) to find the eigenspectrum. Once this is known, we can immediately compute the time dependent solution for the probability distribution using Eq.~(\ref{prob density solution}). This allows us to understand the relaxation process by looking at the slowest eigenvalue. Next, we analyze the Mpemba condition namely $|a^c_2|>|a^h_2|$ (see Sec.~\ref{Distance function and the Mpemba effect}). 
Specifically, this condition is scanned thoroughly to identify the set of initial temperatures for which $|a_2(T)|$ has a non-monotonic behavior with temperature $T$ so that the above-mentioned inequality is satisfied. We provide phase diagrams spanning in the parameter space of $\Delta U$ (will be discussed in the later part) and temperature ratio to underpin the desired regimes for the Mpemba effect. We make an attempt to provide physical reasoning behind all the possible cases.

It is now understood from the discussion in Sec.~\ref{Modulation of the potential and population} that a fully symmetric potential configuration does not lead to the Mpemba effect. In what follows, we first consider an  asymmetric potential configuration. This includes equal widths for the left and right domains and equal heights at the left, center, and right edges of the potential. The only asymmetry is in the form of different depths between the two potential wells. We show in Sec.~\ref{asymmetry not sufficient} that the mere presence of asymmetry in the potential configuration is \textit{not a sufficient} condition to induce the Mpemba effect. To explore further, we take other configurations that have restricted asymmetries.

% As a result, additional asymmetry in several forms might be required in the potential configuration to induce the Mpemba effect. Somewhat surprisingly, we find that no asymmetry is enough to exhibit the Mpemba effect as long as the potential heights at the left, center, and right edges are equal.
  
  This is done by keeping different heights at the left, center, and right edge of the potential. We carefully analyze these different configurations and explore the possibility of the Mpemba effect. The following configurations are of our interest: (a) equal domain widths as discussed in Sec.~\ref{equal domain width}, and (b) unequal domain widths as discussed in Sec.~\ref{unequal domain width}. For both cases (a) and (b), we explore the different possible configurations by varying the depths of the potential wells.

%{\color{red}In what follows, we look at different asymmetries and explore the occurrences of the Mpemba effect. The following configurations are of our interest:  (a)  different widths for the left and right domains of the potential, (b) same domain widths but different positions of the potential wells about the origin, (c) different depths of the potential wells, and (d) for different heights for the left, centre and right edges of the potential  (see Fig. \ref{potential shape}).}

\subsection{\label{asymmetry not sufficient}Asymmetry is not a sufficient criterion}
\begin{figure}
\centering
\includegraphics[width=0.8\columnwidth]{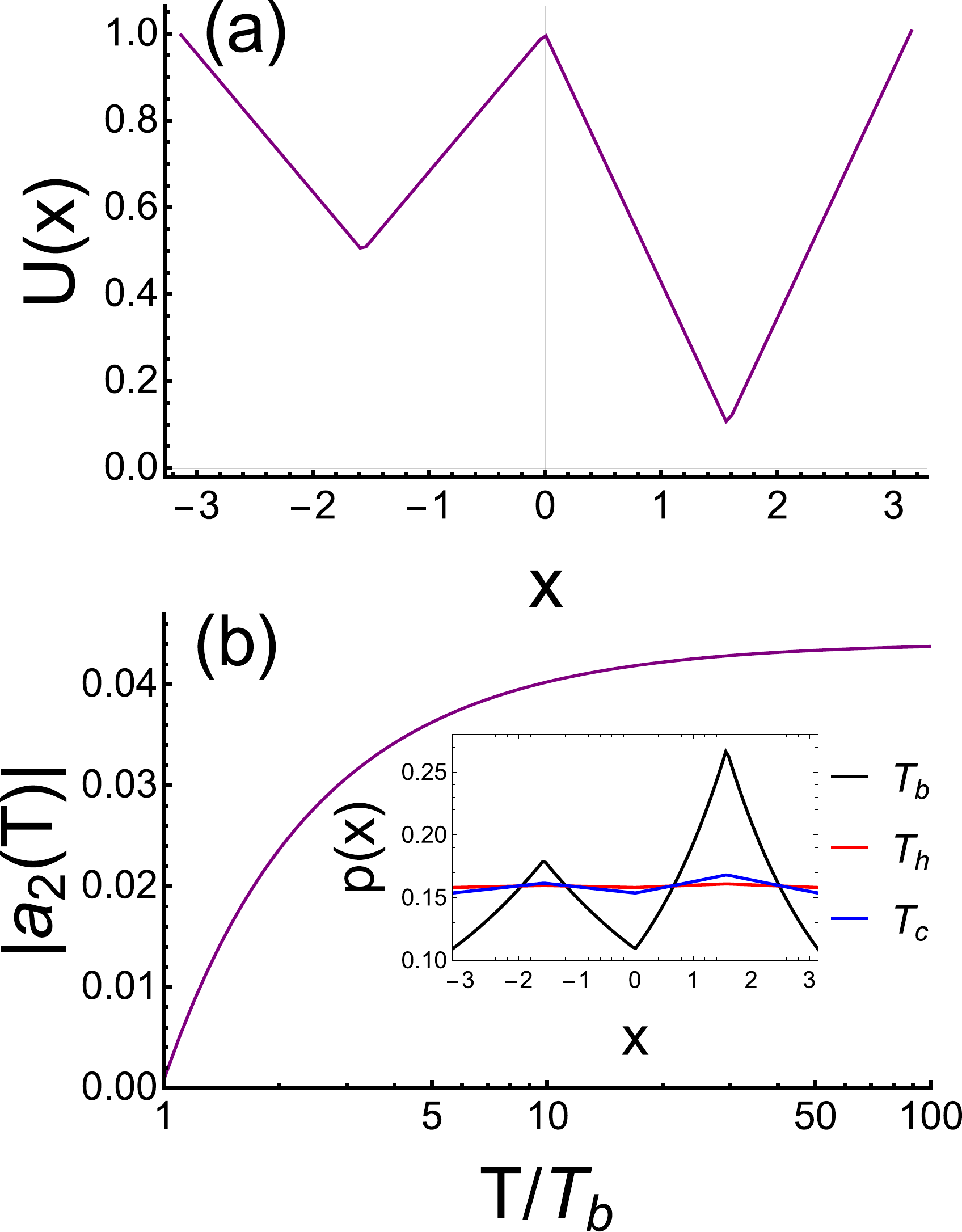}
\caption{Illustration of the absence of the Mpemba effect in an asymmetric double well potential indicating that \textit{asymmetry is not sufficient}.  (a) Asymmetric shape of the potential with different depths for the left and right wells while keeping all the other parameters of the potential symmetric about the origin. 
 The potential heights at its left, center and right edges are equal and so are the positions of the two wells about the origin. The shape of the potential corresponds to the choice of the parameters $x_{max}=x_{min}=\pi$, $\alpha=\beta=0.5$, $k_1=k_2=0.32$ and $k_3=k_4=0.57$. (b) Monotonic evolution of $|a_2(T)|$ with $T$ showing the absence of the Mpemba effect. Inset: Initial population distribution of the confined Brownian particle for the chosen temperatures $T_h=50 T_b$ (red) and $T_c=10 T_b$ (blue) showing almost similarly distributed populations across the potential landscape. The final equilibrium distribution (black) corresponds to bath temperature $T_b=1$.
 }\label{fig combined same height}
\end{figure}

We start by showing that the asymmetry is not a sufficient condition for the existence of the Mpemba effect. As an example, we consider the  case where the asymmetry is only in terms of different depths of the potential wells while keeping everything else symmetric, as shown in Fig.~\ref{fig combined same height}(a). The heights of the left, centre and right edges of the potential well are equal. Moreover, the potential minima are also situated symmetrically about the origin and at the centre of their respective domains. For this case, there is no Mpemba effect since $|a_2(T)|$ increases monotonically with $T$ [see Fig.~\ref{fig combined same height}(b)]. As discussed in Sec.~\ref{Modulation of the potential and population}, the absence of the Mpemba effect can be explained based on the similar nature of the initial population distribution of the hot and the cold system for a particular choice of $T_h$ and $T_c$ respectively, as shown in the inset of Fig.~\ref{fig combined same height}(b), leading to similar relaxations for both the systems.

%As the population of the initially hot and the cold system are similarly distributed in the two wells of the potential landscape, both the systems behave almost identically during the relaxation. 

Hence, one would anticipate that additional asymmetries might be required in the potential configuration to induce the Mpemba effect. However, we find that as long as the potential heights at the left, center, and right edges are equal, there is no Mpemba effect. In what follows, further asymmetric configurations are explored
by considering the cases of equal and unequal domain widths and also varying the depths between the two wells of the potential while satisfying the necessary condition that the heights at the left, center, and right edges of the potential are different.

%cases of equal and unequal domain widths and identify the different configurations leading to the Mpemba effect by also varying the depths between the two wells of the potential while satisfying the necessary condition that the heights at the left, center, and right edges of the potential are different. 

%When the heights at the left, centre and right edge of the potential are different,  we will illustrate several examples where the Mpemba effect exists for the case of a double well potential. Among the various possible asymmetric configurations, we focus on two cases: potential landscape with equal and  unequal domain widths.

\subsection{\label{equal domain width}Equal domain widths}
\begin{figure}
\centering
\includegraphics[width=0.8\columnwidth]{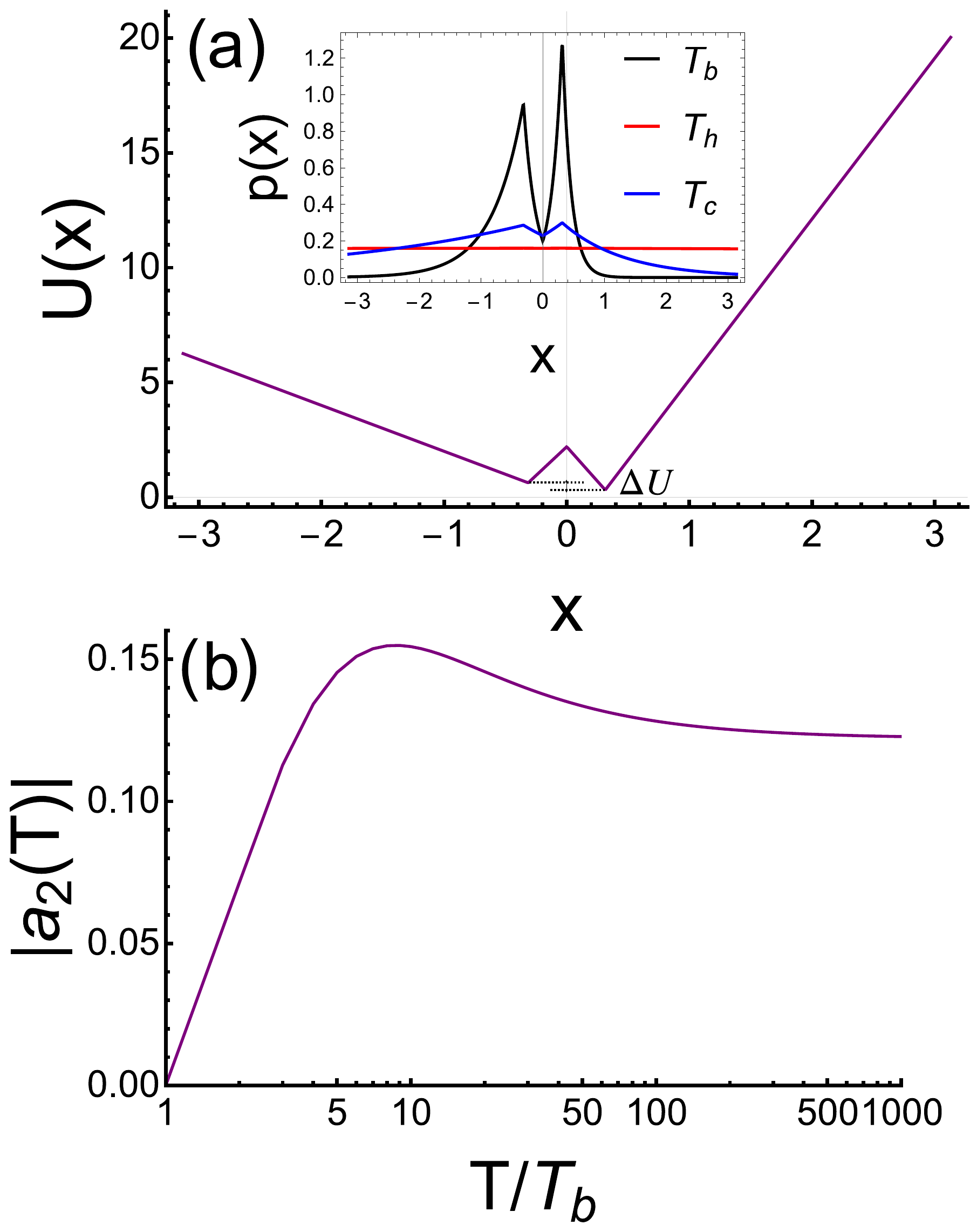}
\caption{Illustration of the Mpemba effect in a confined double well potential with equal domain widths.  (a) Shape of the potential with equal domain widths ($3$ units) about the origin. The asymmetry in the potential configuration is introduced through the choice of different slopes in separate domains. The configuration of the potential is determined by the choice of the parameters: $x_{max}=x_{min}=\pi$, $\alpha=\beta=0.1$, $k_1=2$, $k_2=5$, $k_3=6$ and $k_4=7$. Inset: Initial population distribution of the confined Brownian particle for the chosen temperatures $T_h=1000 T_b$ (red) and $T_c=7 T_b$ (blue) such that $|a^h_2|<|a^c_2|$. This shows a significant population distribution for $T_c$ around the metastable state compared to that (which is almost flat) of $T_h$. The final equilibrium distribution (black) corresponds to bath temperature $T_b=1$. (b) Non-monotonic evolution of $|a_2(T)|$ with $T$ clearly indicates the presence of the Mpemba effect in this set-up.}\label{fig combined symmetric}
\end{figure}

We first examine the configurations of the potential with equal widths for the left and right domains. The boundaries of the well are situated at $x_{min}=x_{max}$ with the position of the two wells equidistant from the origin at $x=-\alpha x_{min}$ and $x=\beta x_{max}$ with $\alpha=\beta$. The various asymmetries in the configuration of the potential are introduced through the choice of the slopes $k_1$, $k_2$, $k_3$ and $k_4$ for the different domains of the confined potential. The shape of the potential with a specific choice of parameters is shown in Fig.~\ref{fig combined symmetric}(a).

The existence of the Mpemba effect for this particular configuration of the potential is evident from  the non-monotonic behavior of the coefficient $|a_2(T)|$ with $T$ as shown in Fig.~\ref{fig combined symmetric}(b).  
%Figure~\ref{fig combined symmetric}(b) shows that the coefficient $|a_2(T)|$ exhibits non-monotonic behavior with $T$ for a particular choice of parameters for this potential configuration. It satisfies the criteria for the Mpemba effect, i.e., $|a^c_2|>|a^h_2|$ (see discussion in Sec.~\ref{Distance function and the Mpemba effect}) for two different temperatures $T_c$ and $T_h$ corresponding to the cold and the hot system respectively.  
The existence of the Mpemba effect for this configuration of the potential is also qualitatively evident in terms of the population distribution of the particle as shown in the inset of Fig.~\ref{fig combined symmetric}(a) for $T_h=1000 T_b$ and $T_c=7 T_b$ which satisfy the criteria for the Mpemba effect. The population distribution of the initially cold system is localised at the intermediate potential well, thus experiencing a metastable state which leads to its slower relaxation towards the final equilibrium. On the other hand, the initially hot system has uniform distribution across the potential landscape and undergoes faster relaxation to the final equilibrium distribution.

%The population distribution of the  particle at $T_c$ is localised at the intermediate potential well when compared to the flat distribution at $T_h$. Thus, the colder system is found to take more time to approach the final equilibrium distribution due to the hindrance caused by the initial localisation at the intermediate well which acts as a metastable state while the initially hot system can relax faster to the final equilibrium distribution as it does not experience such barriers.

\begin{figure}
\centering
\includegraphics[width= \columnwidth]{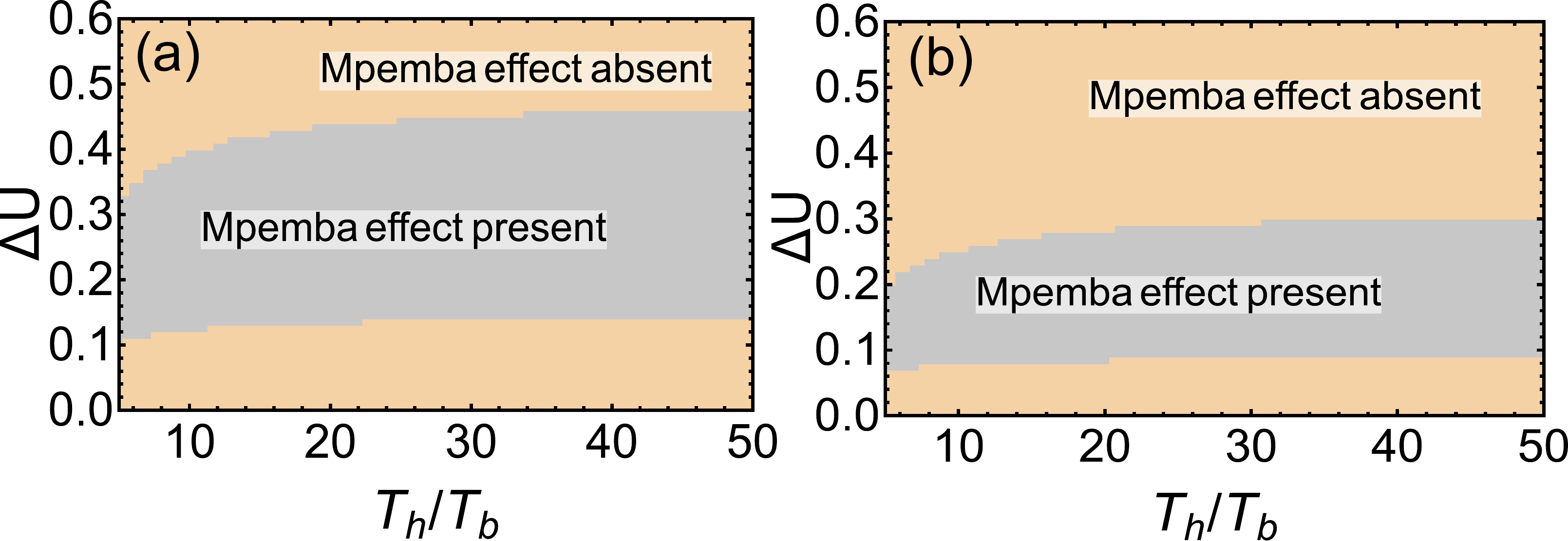} 
\caption{\label{phase diagram symmetric}$\Delta U$-$(T_h/T_b)$ phase diagram illustrating the region of the Mpemba effect in the case of double well potential with equal domain widths. 
% $\Delta U$ denotes the potential difference in the depths of its two wells and $T_h$ is the temperature of the initially hot system measured with respect to the bath temperature $T_b$. 
The phase diagram is obtained by varying the depth of the right well of the potential while keeping the depth of its left well fixed and by changing the temperature ratio. The phase space is partitioned into two domains: one where the Mpemba effect is present corresponding to the criteria $|a^h_2|<|a^c_2|$, and other complementary region. The phase diagrams correspond to different choices of the position of the potential wells which are symmetric about the origin and are determined by: (a) $\alpha=\beta=0.1$, (b) $\alpha=\beta=0.5$.}
\end{figure}

We next explore the phase space that shows the Mpemba effect, for this form of potential configuration in terms of various asymmetries. It is done by varying the depths between the two wells of the potential landscape $\Delta U$ [see Fig.~\ref{fig combined symmetric}(a)] as a function of the temperature of the initially hot system $T_h$ while keeping the temperature of the initially cold system fixed at $T_c=4 T_b$. Changing the depths of the two wells is equivalent to making choices for different possibilities of the slopes $k_1$, $k_2$, $k_3$, and $k_4$.
%We study the phase space of the potential landscape with equal domain widths that shows the Mpemba effect by varying the depths between the two wells of the potential landscape $\Delta U$ [see Fig.~\ref{fig combined symmetric}(a)] as a function of the temperature of the initially hot system $T_h$ while keeping the temperature of the initially cold system fixed at $T_c=4 T_b$. Figures~\ref{phase diagram symmetric}(a) and (b) illustrate the phase diagrams  in the $\Delta U$-$(T_h/T_b)$ plane for two different choices of positions for the center of the potential minima respectively while keeping the widths of the left and the right domains equal and for a fixed choice of the heights for the left, center and right edge of the potential.
Figures~\ref{phase diagram symmetric}(a) and (b) illustrate the phase diagrams of the possible asymmetries in the potential configuration leading to the Mpemba effect, in the $\Delta U$-$(T_h/T_b)$ plane for two different choices of positions for the potential minima although symmetrically placed about the origin.

\subsection{\label{unequal domain width}Unequal domain widths}
We now consider the potential configurations with unequal domain widths and explore the phase space of various possible asymmetries that might demonstrate the Mpemba effect. This is motivated from Ref.~\cite{kumar2020exponentially} where potential with unequal domain widths was considered in order to study the Mpemba effect. 
\begin{figure}
\centering
\includegraphics[width=0.8\columnwidth]{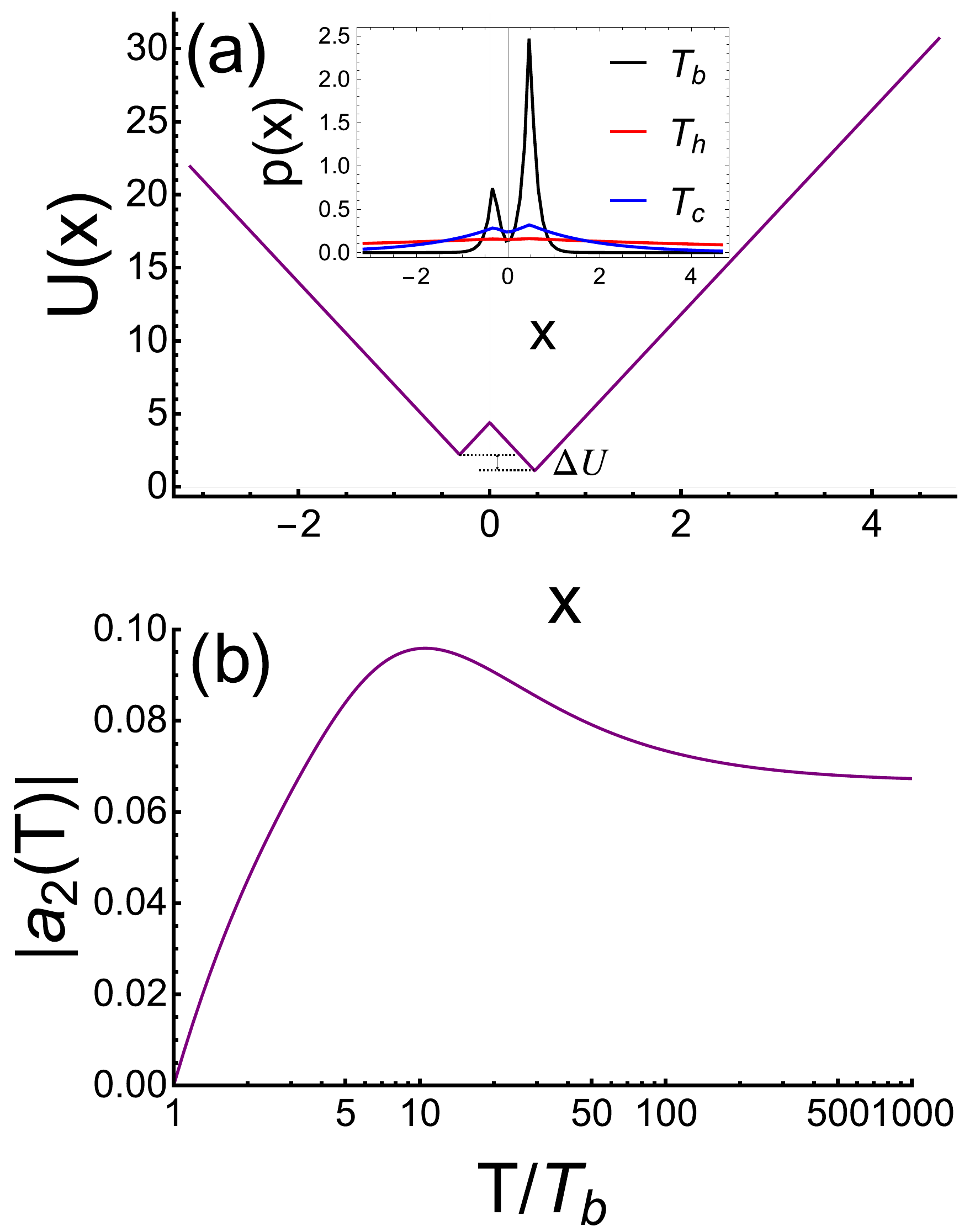}
\caption{Illustration of the Mpemba effect in a confined double well potential with unequal domain widths while keeping every other parameters symmetric about the origin. (a) Shape of the potential with unequal domain widths ($x_{max}\neq x_{min}$) about the origin. The position of the two wells of the potential and its various slopes are kept equal. The configuration of the potential is determined by the choice of the parameters: $x_{min}=\pi$, $x_{max}=1.5 \pi$, $\alpha=\beta=0.1$ and $k_1=k_2=k_3=k_4=7$. Inset: Initial population distribution of the confined Brownian particle for the chosen temperatures $T_h=50 T_b$ (red) and $T_c=10 T_b$ (blue) such that $|a^h_2|<|a^c_2|$. This shows a significant population distribution around the metastable state for $T_c$ compared to the same for 
$T_h$. The final equilibrium distribution (black) corresponds to the bath temperature $T_b=1$. (b) Non-monotonic evolution of $|a_2(T)|$ with $T$ confirms the presence of the Mpemba effect in this set-up.}\label{fig combined asymmetric diff slopes}
\end{figure}

The unequal domain widths of the potential configuration correspond to the positions of its boundaries situated at $x_{min}$ and $x_{max}$ respectively with the magnitudes $x_{max}\neq x_{min}$. The position of the two wells are equidistant from the origin at $x=-\alpha x_{min}$ and $x=\beta x_{max}$ with $\alpha=\beta$. For the simplicity of our analysis, the magnitude of the slopes $k_1$, $k_2$, $k_3$, and $k_4$ are kept equal for different domains. Thus, the only asymmetry in the potential is introduced through the choice of different domain widths of the confined potential, and one such configuration with a particular choice of parameters is shown in Fig.~\ref{fig combined asymmetric diff slopes}(a).

%For this configuration, we study the behavior of the coefficient $|a_2(T)|$ with temperature, $T$ as shown in Fig.~\ref{fig combined asymmetric diff slopes}(b). 

The non-monotonic behavior of the coefficient $|a_2(T)|$ with $T$ as shown in Fig.~\ref{fig combined asymmetric diff slopes}(b) illustrates the existence of the Mpemba effect for this configuration of the potential. We consider one such pair of temperatures $T_h=50 T_b$ and $T_c=10 T_b$ for the hot and cold systems respectively that satisfy the criteria $|a^c_2|>|a^h_2|$ and study the nature of the population distribution of the particle for the particular case as shown in the inset of Fig.~\ref{fig combined asymmetric diff slopes}(a). Here too, the cold system exhibits localisation of its population distribution in the local minima leading to slower relaxation towards the final equilibrium compared to the hot system.
%We find similar localisation of the population distribution for the cold system in the local minima of the potential landscape as discussed earlier for the case of equal domain width of the potential which leads to the Mpemba effect.
%In terms of the population distribution of the particle for the hot and cold systems prepared at the temperatures $T_h=50 T_b$ and $T_c=10 T_b$ respectively
%Since, $|a_2(T)|$ show non-monotonic behavior with  $T$ as shown in Fig.~\ref{fig combined asymmetric diff slopes}(b), there exists the Mpemba effect for any two temperatures $T_h$ and $T_c$ that satisfy the criteria $|a^c_2|>|a^h_2|$. We consider one such pair of temperatures $T_h=50 T_b$ and $T_c=10 T_b$ for the initially hot and cold systems respectively that satisfy the criteria $|a^c_2|>|a^h_2|$ and study the nature of the population distribution of the particle for the particular case as shown in the inset of Fig.~\ref{fig combined asymmetric diff slopes}(a). We find similar localisation of the population distribution for the initially colder system in the local minima of the potential landscape as discussed earlier for the case of equal domain width of the potential which leads to the Mpemba effect. 

\begin{figure}
\centering
\includegraphics[width= \columnwidth]{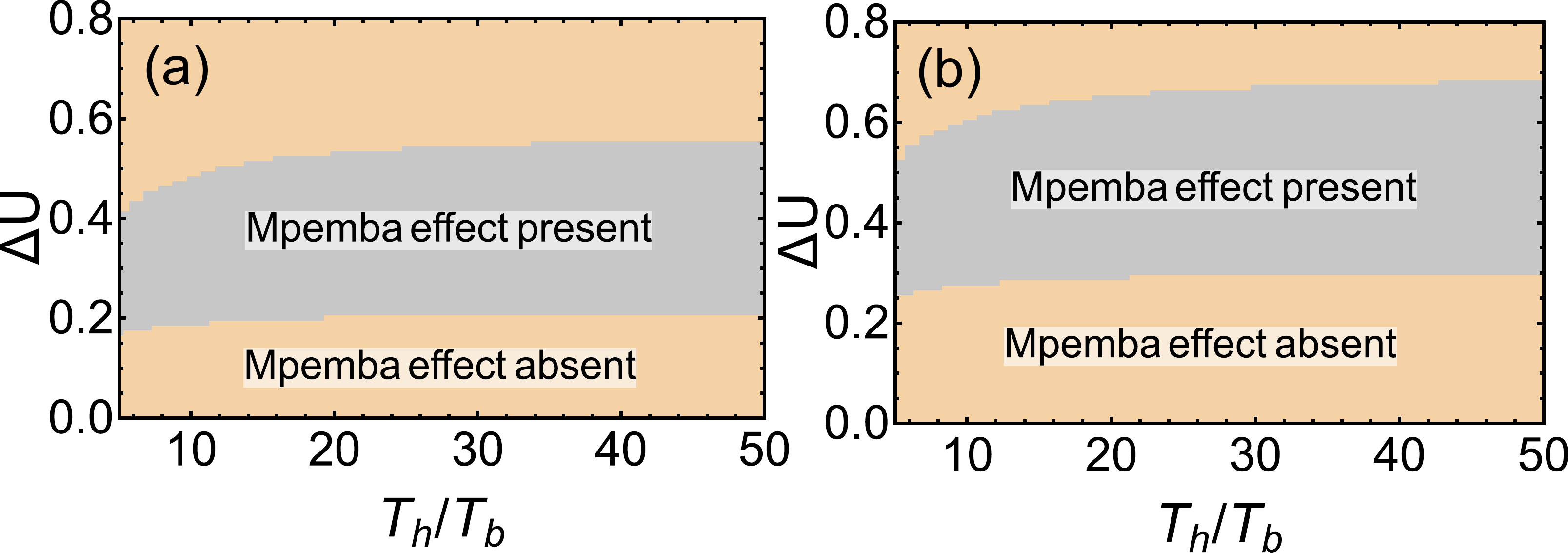} 
\caption{\label{phase diagram asymmetric}$\Delta U$-$(T_h/T_b)$ phase diagram illustrating the region of the Mpemba effect for the case of double well potential 
 with unequal domain widths. The phase diagram is constructed in the similar manner as in Fig. (\ref{phase diagram symmetric}). %obtained by varying the depth of the right well of the potential while keeping the depth of its left well fixed. The phase space of the potential configuration where the Mpemba effect is present for the different temperatures of the initially hot system corresponds to the criteria $|a^h_2|<|a^c_2|$ for a fixed temperature 
The phase diagrams in the left- and right- panel correspond to different choices of the widths for the right domain of the potential : (a) $x_{max}=1.2 \pi$, (b) $x_{max}=1.5 \pi$ keeping the width of its left domain fixed at $x_{min}=\pi$. The position of the two wells are determined by the parameters $\alpha=\beta=0.1$ for  both the cases.}
\end{figure}

We now explore the phase space of possible asymmetries that leads to the Mpemba effect, for this form of potential configuration. Here, the asymmetries are introduced in terms of the choices of different slopes and different widths for the left and right domains. We explore the phase space by varying the depths between the two wells of the potential landscape $\Delta U$ [see Fig~\ref{fig combined asymmetric diff slopes}(a)] as a function of the temperature of the initially hot system $T_h$ while keeping the temperature of the initially cold system fixed at $T_c=4 T_b$. Note that the variation of the two well depths is equivalent to making different choices for the slopes of the potential. We perform this exercise for two different choices of widths for the right domain respectively keeping the width of the left domain fixed as illustrated in Figs.~\ref{phase diagram asymmetric} (a) and (b) respectively. As mentioned earlier, the phase diagrams allow us to provide a comprehensive picture in terms of the parameters that are pertinent to the Mpemba effect.

%We also study the phase space of the potential landscape with unequal domain widths that shows the Mpemba effect by varying the depths between the two wells of the potential landscape $\Delta U$ [see Fig~\ref{fig combined asymmetric diff slopes}(a)] as a function of the temperature of the initially hot system $T_h$ while keeping the temperature of the initially cold system fixed at $T_c=4 T_b$. Figures~\ref{phase diagram asymmetric} (a) and (b) illustrate the phase diagrams  in the $\Delta U$-$(T_h/T_b)$ plane for two different choices of widths for the right domain respectively while keeping the width of the left domain fixed and for a fixed choice of the heights for the left, center and right edge of the potential.

\section{\label{Illustration of the Mpemba effect for single well}Mpemba effect without a metastable minimum}

In this section, we show that the presence of metastable states is not necessary for the existence of the Mpemba effect. We demonstrate this by configuring the potential with no metastable state.  In a recent study, the Mpemba effect was shown  for a piece-wise constant potential  where the local stability of double well potential is replaced by neutral stability~\cite{Walker_2021}. Likewise, we construct potential configurations with no metastable states and yet demonstrate the possibility of observing the Mpemba effect. In short, such an analysis would rationalize the claim that neither metastability nor neutral stability are necessary for the Mpemba effect.

\begin{figure}
\centering
\includegraphics[width=\columnwidth]{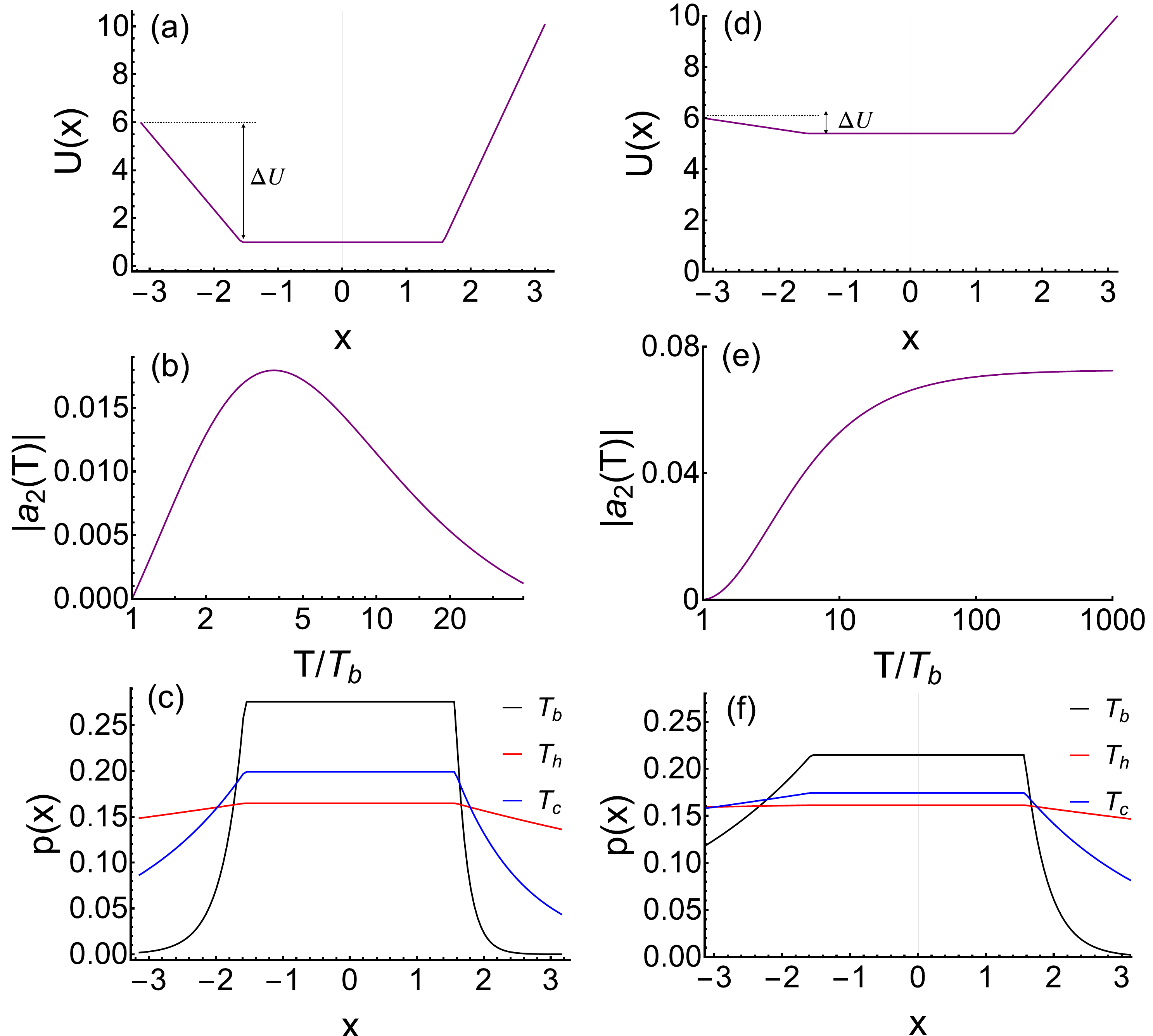}
\caption{Illustration of the Mpemba effect in a confined single well potential with no metastable state. (a) Shape of the single well potential is determined by the choice of the parameters: $x_{max}=x_{min}=\pi$, $\alpha=\beta=0.5$, $k_2=k_3=0$, $k_1=3.18$ and $k_4=5.73$. (b) Non-monotonic evolution of $|a_2(T)|$ with $T$ showing the presence of the Mpemba effect.  (c) Initial population distribution of the confined Brownian particle for the chosen temperatures $T_h=48 T_b$ (red) and $T_c=6 T_b$ (blue). Here, $|a^h_2|<|a^c_2|$ which shows a significant difference in the population density at the minimum of the potential well. The final equilibrium distribution (black) corresponds to bath temperature $T_b=1$. The Mpemba effect disappears for the above configuration of the potential if the depth of the potential well is decreased as shown in (d). The modified potential configuration corresponds to a change in the slopes to $k_1=0.38$ and $k_4=2.93$ while keeping the other parameters same as in the earlier case. (e) Monotonic evolution of $|a_2(T)|$ with $T$ shows the absence of the Mpemba effect for the modified configuration. (f) Initial population distribution of the confined Brownian particle for the same pair of temperatures $T_h=48 T_b$ (red) and $T_c=6 T_b$ (blue) show nearly similar population distribution at the minimum of the potential well.}\label{fig combined no metastable state}
\end{figure}

Let us consider the single well potential with two linear slopes at the edges and with fixed magnitude in between $x=-\alpha x_{min}$ to $x=\beta x_{max}$ (see Fig.~\ref{fig combined no metastable state}), where $x_{min}$ and $x_{max}$ are the boundaries where the potential goes to infinity. We find that the minimal criterion to observe the Mpemba effect in this configuration is to introduce an asymmetry in the form of different heights for the left and right edge of the potential landscape with $\alpha=\beta < 1$.

One such configuration of the potential is shown in Fig.~\ref{fig combined no metastable state}(a). The existence of the Mpemba effect for this configuration is illustrated through the non-monotonic behavior of the coefficient $|a_2(T)|$ with temperature $T$ -- see Fig.~\ref{fig combined no metastable state}(b).
%We illustrate the existence of the Mpemba effect for the potential configuration shown in Fig.~\ref{fig combined no metastable state}(a). For this configuration, the coefficient $|a_2(T)|$ increases non-monotonically with the increase in temperature, $T$ as shown in Fig.~\ref{fig combined no metastable state}(b) satisfying the criteria for the Mpemba effect.
%We illustrate the existence of the Mpemba effect for a given configuration of the single well potential landscape as shown in Fig.~\ref{fig combined no metastable state}(a). For this configuration, we study the behavior of the coefficient $|a_2(T)|$ with temperature, $T$ as shown in Fig.~\ref{fig combined no metastable state}(b). Since, $|a_2(T)|$ show non-monotonic behavior with  $T$, there exists Mpemba effect for any two temperatures $T_h$ and $T_c$ that satisfy the criteria $|a^c_2|>|a^h_2|$. 
We consider one such pair of temperatures $T_h=48 T_b$ and $T_c=6 T_b$ for the initially hot and cold systems respectively that satisfy the criteria $|a^c_2|>|a^h_2|$ and study the nature of the population distribution of the Brownian particle for the particular case as shown in Fig.~\ref{fig combined no metastable state}(c).

In the case of the double well potential configuration, the presence of a metastable state plays an important role in the existence of the Mpemba effect.
%As discussed earlier for the case of the double well configuration of the potential landscape, the presence of a metastable state  causes hindrance in the redistribution of the population of the particle corresponding to the final equilibrium from an initially cold state which qualitatively describes the Mpemba effect. 
Clearly, in this case, there is no delay in the redistribution of the populations to the final equilibrium distribution starting from two different temperatures due to the  absence of any metastable state. However, the existence of the Mpemba effect for this configuration shows that there is a trade-off between the initial population density and kinetic energy of the particle in the redistribution process to the final equilibrium as evident from the population statistics near the edge of the potential landscape [see Fig.~\ref{fig combined no metastable state}(c)]. 

Although the initially hot system has more population of the particles near the edge of the potential to redistribute than the same of the initially cold system, the higher kinetic energy of the hot system dominates during the relaxation process for the given configuration of the potential landscape, leading to a faster relaxation of the hot system than the cold one and hence the Mpemba effect is observed.

However, keeping the same configuration of the potential landscape and same temperatures for the initial hot and the cold system, we find that the anomalous relaxation disappears as the depth of the potential minimum is decreased as is illustrated in Fig.~\ref{fig combined no metastable state}(d) and (e). It is 
evident from the monotonically increasing nature of the coefficient $|a_2(T)|$ with temperature, $T$ [see Fig.~\ref{fig combined no metastable state}(e)] that there is no Mpemba effect in this case. The population distribution of the particle for this configuration of the potential is shown in  Fig.~\ref{fig combined no metastable state}(f). A qualitative argument can be given based on the trade-off between the initial population density and kinetic energy of the particles present at the edges of the potential landscape. The presence of a smaller population for the initially cold system at the edges (which would eventually redistribute to the potential minimum) dominates in the relaxation process to the final equilibrium. Naturally, one would expect that the initially cold system will approach the final equilibrium faster than the initially hot system discarding the possibility of a Mpemba effect.

%evident from the behavior of the coefficient $|a_2(T)|$ with temperature, $T$ [see Fig.~\ref{fig combined no metastable state}(e)]. Since, $|a_2(T)|$ increases monotonically with  $T$,  there is no Mpemba effect [see discussion in Sec.~\ref{Distance function and the Mpemba effect}] for this configuration of the potential landscape.  

\begin{figure}
\centering
\includegraphics[width= 0.8\columnwidth]{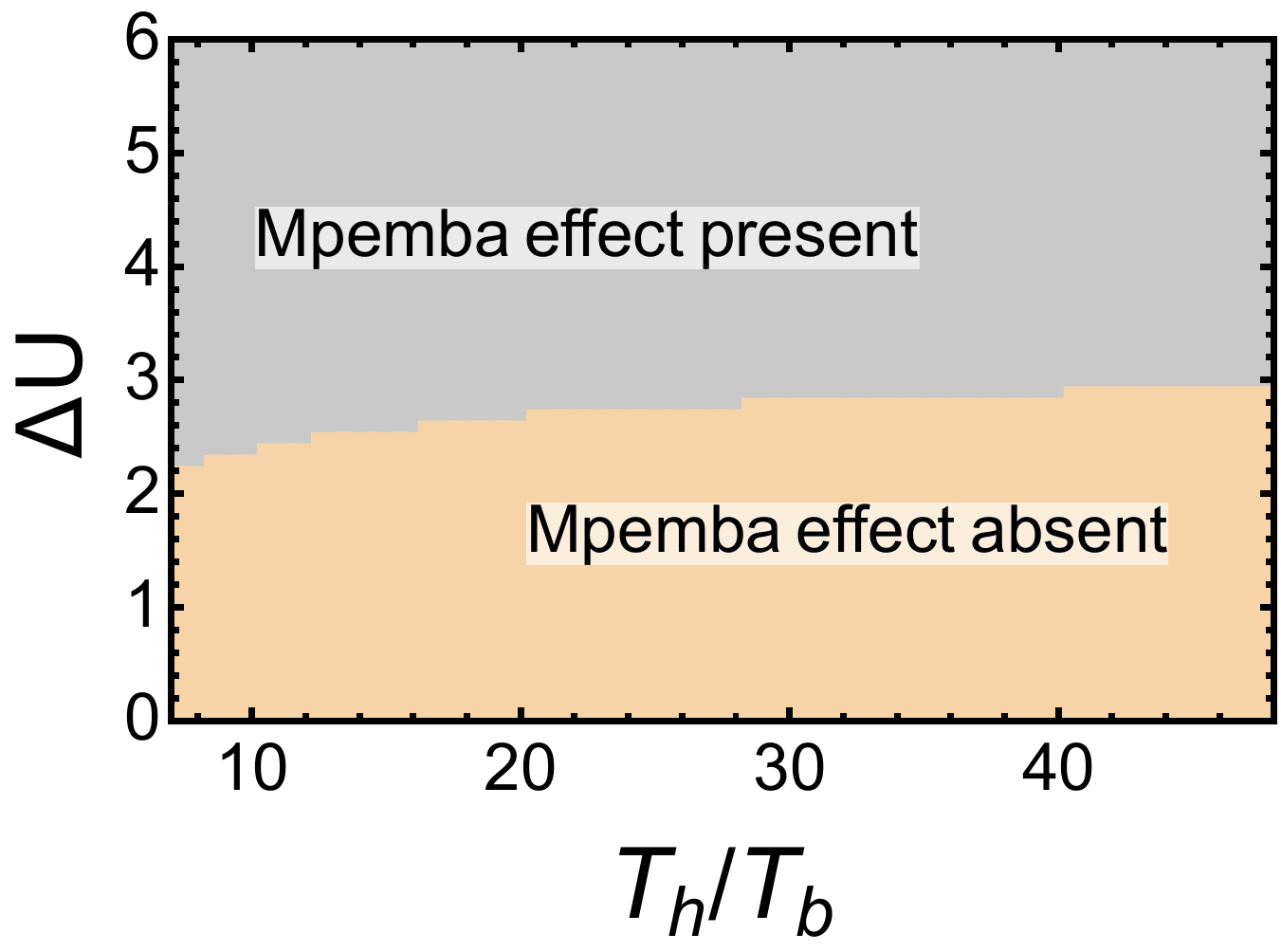} 
\caption{\label{phase diagram single well}$\Delta U$-$(T_h/T_b)$ phase diagram illustrating the region of the Mpemba effect for the case of a single well potential. The phase diagram is obtained as before by varying the depth of the potential minimum \& the temperature ratio. Here, there are two distinct regions and the criterion $|a^h_2|<|a^c_2|$ marks the one where the Mpemba is observed. Here, $T_c=4T_b$ as before and the other parameters determining the configuration of the potential are:            $\alpha=\beta=0.5$.}
\end{figure}

Finally, we explore the phase space of the single well potential landscape with the temperature ratio. We vary the minimum or the depth of the potential landscape $\Delta U$ measured with respect to the potential height at the left edge [see Figs.~\ref{fig combined no metastable state}(a) and (d)] as a function of the temperature of the initially hot system $T_h$ while keeping the temperature of the initially cold system fixed at $T_c=4 T_b$. Figure~\ref{phase diagram single well} illustrates the phase diagram  in the $\Delta U$-$(T_h/T_b)$ plane for a fixed choice of the heights for the left and right edge of the potential.

\section{\label{Conclusion}Conclusion }

In summary, we have theoretically studied the Mpemba effect in a system of an overdamped particle trapped in an external potential motivated by a similar experimental set-up \cite{kumar2020exponentially}. The potential is generically piece-wise linear but double-welled, and moreover we can maneuver it to give various shapes. One can exactly solve this model analytically to obtain the eigenspectrum decomposition  of the corresponding Fokker-Planck equation. This allows us to provide a comprehensive study of the Mpemba effect spanning a wide panorama of  physical scenarios.
% The potential closely mimics the experimentally studied~\cite{kumar2020exponentially} potential which is double-welled, but is quadratic instead of linear. 

As noted earlier in \cite{kumar2020exponentially} and in other works that symmetric potentials are not expected to exhibit the Mpemba effect. For symmetric potentials, $a_2$ is exactly zero and higher order coefficients become important. By explicit calculation, absence of the Mpemba effect was noted in piece-wise constant potential as well as the pure harmonic potential~\cite{Walker_2021}. We expect the same to hold for symmetric piece-wise linear potentials. For the class of symmetric potentials that we explored, we did not find any exceptions. Asymmetry was introduced in the experiment in Ref.~\cite{kumar2020exponentially} through different domain widths for the two minima. We also demonstrate the existence of the Mpemba effect when the two widths are unequal. Through counterexamples, we also show that unequal domain widths are neither a necessary nor a sufficient condition for the Mpemba effect to be present. We also show that the Mpemba effect can be realized for equal domain widths but for other asymmetries in the potential. In particular, we find that the Mpemba effect is easily realizable when the heights of the potential at the left, center and right edges are different. This is a notable feature of our work.

Concluding, the Mpemba effect in Langevin systems is usually depicted in terms of the ruggedness in the energy landscape where the particles diffuse. The relaxation of the colder system to the lowest energy state is usually hindered by the presence of metastable states while the hotter system does not experience (i.e., can overlook) the metastable states due to its higher energy and thus can relax to the lowest energy state faster than the colder system. In this paper, we revisit this physical picture for a variety of different cases.
% that is studied and the explanation is given in terms of the concentration of the population of the particle as a result of modulation of the external potential. 
Generically it is understood that the larger energy barrier leads to a significant amount of population concentration at the intermediate energy well (or the metastable state) for the initially colder system as compared to the initially hotter system. This leads to 
the consensus that metastability might be necessary for the Mpemba effect. We benchmark this rationale within our exactly solvable model.
However and in stark contrast, we also show that metastable states are not necessary for Mpemba effect by demonstrating the effect in a potential with \textit{no metastable states}, questioning the current qualitative understanding. This result also improves on the result in Ref.~\cite{Walker_2021}, where for piece-wise constant potentials, metastability was replaced by neutral stability. Taken together these new observations, we believe that our work offers a significant aid to the current understanding of the Mpemba effect in Langevin systems.
% Exploring the Mpemba effect further in single-well potential set-up is an ongoing work. 
Finally, it is within our understanding that there is a subtle interplay between the initial population (manifesting the energy landscape) and the  hopping frequency for particle rearrangement (which crucially depends on the temperature), however a rigorous quantification is yet to be made. This is a key aspect that requires further investigations.

\section{Acknowledgement}
Arnab Pal gratefully acknowledges research support from the DST-SERB Start-up Research Grant Number SRG/2022/000080.

\appendix
\onecolumngrid
\section{ Finding the constants $C_n$ and $D_n$ \label{appendix 1}}
In this appendix, we solve for the constants $C_n$ and $D_n$ in terms of $B_n$ and $H_n$.
Solving for $C_n$ and $D_n$ in terms of $B_n$ using Eqs.~(\ref{coeff 1}) and (\ref{coeff 2}), we obtain:
\begin{align}
C_n=&\frac{B_n}{2 m_2} \Big[-\Big((k_1 + k_2-2 m_1 \nu_{1n})\cos(m_2 \alpha x_{min}) + 2 m_2 \sin(m_2 \alpha x_{min}) \Big) \cos(m_1 \alpha x_{min}) \nonumber \\
& + \Big(\big((k_1 + k_2)\nu_{1n}+2 m_1 \big)\cos(m_2 \alpha x_{min}) + 2 m_2 \nu_{1n} \sin(m_2 \alpha x_{min}) \Big) \sin(m_1 \alpha x_{min})  \Big], \label{C1}\\
D_n=&\frac{B_n}{2 m_2} \cos(m_2 \alpha x_{min}) \Big[\Big(-2m_2 \nu_{1n} + (2m_1+(k_1+k_2)\nu_{1n})\tan(m_2\alpha x_{min}) \Big)\sin(m_1 \alpha x_{min}) \nonumber \\
&+ \Big(2 m_2-(k_1+k_2-2m_1\nu_{1n})\tan(m_2 \alpha x_{min}) \Big)\cos(m_1 \alpha x_{min}) \Big]. \label{D1}
\end{align}

Solving for $C_n$ and $D_n$ in terms of $H_n$ using Eqs.~(\ref{coeff 3}) and (\ref{coeff 4}), we obtain:
\begin{align}
C_n&=\frac{-8m^2_3 H_{n}}{m_2 [k^2_2-k^2_3+8m^2_3+(k^2_3-k^2_2)\cos(2m_3 \beta x_{max}) ]} \nonumber \\
&\times\Bigg[\Big[\frac{k_2}{2} \cos(m_3\beta x_{max})-\big(m_3+\frac{k_2(k_2+k_3\big)}{4m_3})\sin(m_3\beta x_{max}) \Big] \Big[\cos(m_4\beta x_{max})-\nu_{4n} \sin(m_4\beta x_{max}) \Big] \nonumber \\
&-\Big[\frac{1}{2}\cos(m_3\beta x_{max}) + \frac{k_2+k_3}{4m_3}\sin(m_3\beta x_{max}) \Big]  \Big[ (k_4-2m_4 \nu_{4n}) \cos(m_4 \beta x_{max})-(2m_4 + \nu_{4n} k_4)\sin(m_4 \beta x_{max}) \Big]\Bigg], \label{C2}\\
D_n&=\frac{4m_3 H_n}{k^2_2-k^2_3+8m^2_3+(k^2_3-k^2_2)\cos(2m_3 \beta x_{max}) } \times \Bigg[2m_3 \cos(m_3\beta x_{max}) \Big[\cos(m_4\beta x_{max})-\nu_{4n} \sin(m_4\beta x_{max}) \Big] \nonumber\\
&\quad+\sin(m_3\beta x_{max})\Big[(2 m_4 + (k_3+k_4)\nu_{4n}) \sin(m_4 \beta x_{max})-(k_3+k_4-2m_4\nu_{4n}) \cos(m_4 \beta x_{max})\Bigg].
 \label{D2}
\end{align}

\twocolumngrid

%\bibliography{ref-mpemba}
%aipnum4-2.bst 2019-01-14 (MD) hand-edited version of apsrev4-1.bst

\end{document}